\documentclass[preprint,12pt]{elsarticle}

\usepackage{epsfig}
\usepackage{amssymb}
\usepackage{amsmath}
\usepackage{bbold}
\usepackage[dvipsnames]{xcolor}
\usepackage[normalem]{ulem}

\newcommand\Rsout{\bgroup\markoverwith{\textcolor{red}{\rule[0.5ex]{2pt}{0.4pt}}}\ULon}
\newcommand\Bsout{\bgroup\markoverwith{\textcolor{blue}{\rule[0.5ex]{2pt}{0.4pt}}}\ULon}

\newcommand{\as}{\ensuremath{\alpha_s}}

\newcommand{\msb}{\ensuremath{\overline{\rm MS}}\ }
\newcommand{\mt}{\ensuremath{\tilde{m}}}

\usepackage[ragged]{footmisc}
\usepackage[compat=1.1.0]{tikz-feynman}
\usetikzlibrary{snakes,arrows,decorations,backgrounds}

\usepackage{tikz}
\usepackage[customcolors]{hf-tikz}
\usepackage{mciteplus}

\usetikzlibrary{arrows,cd,shapes,decorations.pathmorphing,decorations.markings,shadings}
\tikzset{
  big arrow/.style={
    decoration={markings,mark=at position 1 with {\arrow[scale=4,#1]{>}}},
    postaction={decorate},
    shorten >=0.4pt},
  big arrow/.default=blue}

\ExplSyntaxOn
\NewDocumentCommand{\newwhiledo}{m m}
  {
   \bool_while_do:nn { \int_compare_p:n {#1} } { #2 }
  }
\ExplSyntaxOff

\def\picSc{0.7} 
\def\blobSc{0.5} 

\def\lineW{0.5} 

\newcommand\leftArc[2]{
	\pgfmathsetmacro{\np}{#1} 
	\ifthenelse{\equal{\np}{1}}	{
		\def\angle{180}
		\node[circle,line width=\lineW mm,draw,fill=Gray!10] (V\np) at (\angle: 1)[scale=\blobSc] {#2};
	}	{
		\foreach \k in {1,...,\np}
			\def\angle{{ 360 * ( \k / (2*\np+2) ) + 90 }}
			\node[circle,line width=\lineW mm,draw,fill=Gray!10] (V\k) at (\angle: 1)[scale=\blobSc] {#2};
			\draw [line width=\lineW mm]  \foreach \x [remember=\x as \lastx (initially 1)] in {2,...,\np}{(V\lastx) to [bend right] (V\x)};
	}
	\draw [line width=\lineW mm] (V1) to [bend left] (0,1);
	\draw [line width=\lineW mm] (V\np) to [bend right] (0,-1);
}

\newcommand\rightArc[2]{
	\pgfmathsetmacro{\np}{#1} 
	\ifthenelse{\equal{\np}{1}}	{
		\def\angle{0}
		\node[circle,line width=\lineW mm,draw,fill=Gray!10] (V\np) at (\angle: 1)[scale=\blobSc] {#2};
	}	{
		\foreach \k in {1,...,\np}
			\def\angle{{ -1 * 360 * ( \k / (2*\np+2) ) + 90 }}
			\node[circle,line width=\lineW mm,draw,fill=Gray!10] (V\k) at (\angle: 1)[scale=\blobSc] {#2};
			\draw [line width=\lineW mm] \foreach \x [remember=\x as \lastx (initially 1)] in {2,...,\np}{(V\lastx) to [bend left] (V\x)};
	}
	\draw [line width=\lineW mm] (V1) to [bend right] (0,1);
	\draw [line width=\lineW mm] (V\np) to [bend left] (0,-1);
}

\newcommand\doubleGluonLoop[3]{
	\draw[decorate,  style=double, decoration={snake,amplitude=.4mm,segment length=2mm,post length=0mm}, line width=\lineW mm, Black!80,fill=White] (#1,#2) circle (#3);
}
	
\begin{document}

\begin{frontmatter}



\title{Cold Quark Matter: Renormalization group improvement of the perturbative series }


\author{Loïc Fernandez\footnote{Speaker}} 
\affiliation{organization={Helsinki Institute of Physics},
            addressline={P.O Box 64, FI-00014, University of Helsinki},
            city={Helsinki},
            country={Finland}}
         
\author{Jean-Loïc Kneur}
\affiliation{organization={Laboratoires Charles Coulomb (L2C)},
            addressline={UMR 5221, Université de Montpellier},
            city={Montpellier},
            country={France}}            

\begin{abstract}
We discuss recent improvements of the cold and dense QCD pressure owing to an all-order resummation of the soft modes, or to the so-called renormalization group optimized perturbation theory (RGOPT). 
Both approaches show a significant improvement of the residual renormalization scale dependence with respect to the state-of-the-art results 
for the perturbative pressure.
\end{abstract}


\end{frontmatter}



\section{Introduction}

In Quantum Chromodynamics (QCD) 
at finite temperature and/or density, 
in-medium thermodynamic parameters are responsible for the emergence of long-range correlations between gluonic fields, rendering analytical approaches, based on local operators, challenging. Nevertheless, upon identifying and resumming a certain subclass of infrared divergent (IR) diagrams, a finite weak-coupling expansion result can be achieved. 
Three different scales naturally emerge at finite temperature $T$ and/or baryonic chemical potential
($\mu_B$): the hard scale $(T,\, \mu_B)$, the soft scale $(\sqrt{\as}T,\sqrt{\as}\mu)$ and the ultra-soft scale $\as\, T$ associated with the Linde\cite{Linde} nonperturbative problem, starting at ${\cal O}(\alpha_s^3)$. 
At finite temperatures, this identification and classification is well understood through the use of effective field theory (EFT) description such as EQCD\cite{EFTpQCD,EFTpQCDg6}, or alternatively the Hard Thermal Loop (HTL)\cite{HTLbasic} 
EFT. For vanishing baryonic densities and high temperatures, the state-of-the-art thermodynamical calculations\cite{lattice} are held by lattice QCD (LQCD) simulations.
In the opposite region of the phase diagram, with vanishing temperatures and high baryonic chemical potential $\mu_B$, LQCD is plagued by the sign problem for quark fields\cite{sign}, preventing direct application to 
the mid- and high-range $\mu_B$ values. 
On the other hand, the nonperturbative ultrasoft scale problem is not relevant at zero temperature,
and perturbative QCD (pQCD) is reliable for relatively high density (equivalently chemical potential) values, once the IR (soft) sector has been properly addressed. 
However, the soft
scale being responsible for the breakdown of naive perturbative expansion, it requires a resummation, which in turn leads to non-analytical coupling dependencies in thermodynamical quantities. The pioneering calculation by Freedman and McLerran\cite{soft1} of the next-to-next-to-leading order (NNLO) pressure at vanishing temperature and finite chemical potential (for massless quarks, later generalized to the massive case\cite{Kurkela:2009gj}) displayed the emergence of $\as^2 \ln \as$ dependence from the plasmon (ring) resummation of the soft sector.
In a more modern picture, although one cannot ``integrate out'' the hard scale in the same
manner as for $T \ne 0$, the soft and hard scale results can nevertheless be organized neatly\cite{Gorda:2018gpy,Gorda:2021znl,Gorda:2021kme} according to a power counting argument which is more easily understood using the HTL\cite{HTLbasic} EFT.
Tremendous efforts have recently been made towards the completion of the $\rm N^3LO$\cite{Gorda:2018gpy,Gorda:2021znl,Gorda:2021kme,Gorda:2023mkk,Karkkainen:2025nkz,Navarrete:2024zgz}
and there remains now only one missing contribution from the hard sector.
Also recently, upon exploiting the properties of the HTL EFT renormalization group, the $\ln\alpha_s$ dependence has been identified to belong to a specific family of (soft) logarithms linked by the renormalization group, leading to their all-order resummation\cite{Fernandez:2021jfr}.

One reason of the HTL success is to systematically expand around a quasiparticle mass, 
which acts as an IR cutoff. This is reminiscent of approaches also used at zero temperature and density, the Hartree approximation and its generalizations based on interpolated Lagrangians augmented by variational prescriptions aiming to go beyond strictly perturbative expansions. Such
optimized perturbation theories (OPT)\footnote{There is a vast literature on similar approaches under various names: order-dependent mapping (ODM)\cite{OPT_ODM}, OPT\cite{OPT_PMS}, $\delta$-expansion (see, e.g., \cite{OPT_LDE}), variational perturbation theory (VPT)\cite{VPT_Kleinert}, etc.}, or similarly the screened perturbation theory\cite{spt1,spt3} (SPT) in the thermal context,
have found a lot of applications when used in conjunction with the HTL Lagrangian, hence known as Hard Thermal Loop perturbation theory 
(HTLpt)\cite{HTLpt}. 
   More recently, a RG-improved variant of OPT (or SPT), dubbed renormalization group optimized perturbation theory (RGOPT), has been developed\cite{RGOPTals,KNcond}. At finite temperatures, it has been investigated in $\phi^{4}$ theory \cite{rgopt_phi4} up to NNLO \cite{rgopt_phi4_NNLO}, and also at NLO for the QCD pressure\cite{rgopt_hot}, where in both cases it drastically reduces the dependence on the residual renormalization scale compared to the standard weak-coupling expansion, SPT or HTLpt. Moreover, for hot QCD the NLO RGOPT is quite close to lattice results\cite{lattice} (at least down to temperatures moderately above $T_c$). Concerning cold quark matter, the NLO RGOPT\cite{rgopt_cold} also reduces the residual scale dependence, although more moderately than in the $T\ne 0$ case.

In the present communication, we review in cold and dense QCD the resummation of the soft leading logarithms\cite{Fernandez:2021jfr} in the framework of HTL EFT, as well as the recent application of RGOPT at NNLO\cite{RGOPT_ColdDense_NNLO}. 

\section{Resummation of the soft leading logarithms}
\subsection{Cold quark matter at NNLO}
For $N_f$ massless quark flavors with degenerate chemical potential $\mu_u=\mu_d=\mu_s=\mu$, 
the NNLO weak coupling expansion pressure
reads\cite{soft1,pQCDmu3l}
\begin{equation}\label{eq:PQCD_massless}
    \hspace{-0.6cm}\frac{\mathcal{P}^{\rm QCD}_{\rm NNLO}(\mu,\alpha_s(\Lambda))}{\mathcal{P}_{f}}=1-\frac{2}{\pi}\as-\as^2\left(\frac{N_f}{\pi^2}\ln\as-0.874355-16\frac{(11N_c-2N_f)}{3(4\pi)^2}\ln\left(\frac{\Lambda}{\mu}\right)\right).
\end{equation}
 As previously alluded to, the $\ln\as$ appearing at $\as^2$ stems from the IR resummation of the naive perturbative expansion, and these contributions can be systematically organized and evaluated 
 using HTL EFT~\footnote{Note an important distinction between the usage of HTL EFT for cold quark matter
 as described here, and HTLpt: 
the former is used as an EFT framework to organize the IR sector of pQCD. While the latter incorporates extra ingredients and prescriptions aiming to go beyond
standard pQCD, as mentioned above.}. It incorporates resummation of all leading-order Hard Thermal Loops inside effective propagators and vertices. This is encapsulated in an effective Lagrangian first derived by Braaten and Pisarski\cite{HTLbasic} 
\begin{equation}\label{eq:Lagrangien_HTL}
    \mathcal{L}_{\rm EFT}^{\rm LO}\equiv\mathcal{L}_{\rm YM}^{\rm HTL}=-\frac{m_E^2}{2}{\rm Tr}\, \int_{\hat{v}}F^{\alpha\beta}\frac{v_\beta v^\gamma}{\left(v\cdot D \right)^2}F_{\gamma\alpha},
\end{equation}
where $D$ is the covariant derivative in the adjoint representation and $v^\mu=(1,\hat v)$ with $\hat v$ a unit 3-vector.
At leading order in HTL EFT, only the soft sector of QCD is included. At next-to-leading order, contributions from two new operators arise, referred to as the ``mixed'' sector\cite{Gorda:2021znl}, and they describe how the hard sector back-reacts on the soft one and conversely. As for the hard sector, it is not included in the EFT but added order by order via a matching equation (see e.g. \cite{coolQM})
\begin{equation}\label{eq:Matching}
    \mathcal{P}_{\rm QCD}\simeq  \mathcal{P}_{\rm EFT}+\left( \mathcal{P}_{\rm QCD}-\mathcal{P}_{\rm EFT}\right)\big|_{\rm expanded\ in\ m_E} +{\rm higher\ orders}.
\end{equation}
The operator in Eq.(\ref{eq:Lagrangien_HTL}) generates a one-loop resummed gluon diagram 
\begin{eqnarray}\label{eq:PEFT_NLO}
    \mathcal{P^{\rm EFT}_{\rm LO}}= \begin{tikzpicture}[baseline=-\the\dimexpr\fontdimen22\textfont2\relax,scale=1.2*\picSc]
	\doubleGluonLoop{0}{0}{0.5}
 \end{tikzpicture} =&& \hspace{-5mm} \int\frac{d^D P}{(2\pi)^D}\left\{(D-2)\ln\left(P^2+m_E^2\, \Pi_{T}(P)\right)+\ln\left(P^2+m_E^2\, \Pi_{L}(P)\right)\right\}, \nonumber \\
 =&&\frac{(N_c^2-1)\,m_E^4}{\left(8\pi\right)^2}\left(\frac{1}{2\varepsilon}
 -\ln\left(\frac{m_E}{\Lambda}\right) +1.17201\right) .
\end{eqnarray}
Once the Debye mass $m_E$ has been evaluated at leading order,
\begin{equation}\label{eq:mE}
    m_E^2=2\frac{\as}{\pi}\sum_f \mu_f = 2\frac{\as}{\pi}N_f \mu,
\end{equation}
it entails the $\as^2\ln\as$ term that appears in \eqref{eq:PQCD_massless}. Note that within the HTL EFT picture at leading order, both the UV divergence and $\ln(\frac{\Lambda}{\mu})$ terms remnant
in Eq.(\ref{eq:PEFT_NLO}) get canceled upon matching with the hard sector\cite{Gorda:2021znl,Gorda:2023mkk}, Eq.\eqref{eq:Matching}, 
respectively, by a corresponding IR divergence and logarithmic term, then giving the complete NNLO expression \eqref{eq:PQCD_massless}.  
Now in contrast, prior to matching $m_E$ onto Eq.(\ref{eq:mE}), 
 the HTL mass term develops its own anomalous (mass) dimension just like any operator within the EFT: $\gamma_m^g(\as)=\gamma_0^g\, \as+\mathcal{O}(\as^2)$, therefore, modifying the RG operator according to
\begin{equation}\label{eq:RGeqHTL}
    \Lambda\frac{d}{d\Lambda}=\Lambda\frac{\partial}{\partial\Lambda}+\beta^g(\as)\frac{\partial}{\partial\as}-m_E\, \gamma_m^g(\as)\frac{\partial}{\partial m_E},
\end{equation}
where $\beta^g(\as)$ is the pure gauge beta function. 

\begin{figure}[h!]\label{fig:ResumLL}
\centering
\epsfig{file=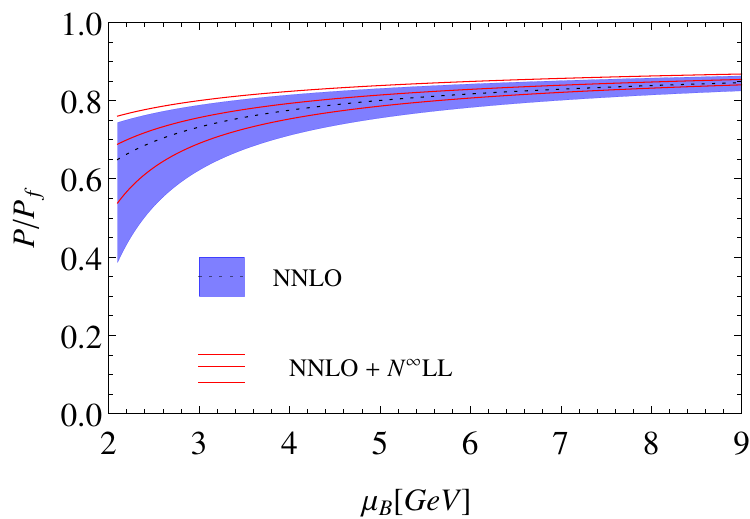,width=0.65\linewidth}
\caption{Plain blue: standard NNLO 
cold and dense QCD pressure $P(\mu_B)$ normalized to the free quark gas, 
with renormalization scale dependence  $\Lambda\in\{\mu,4\mu\}$. Red: the same quantity plus the all-order resummation of soft-leading logarithms.}\label{fig1}
\end{figure}

This apparently simple modification unravels a rich structure that is not accessible by the massless RG in QCD.
If one were to 
use the massless RG operator on the full QCD pressure, i.e. matching $m_E$ to its value \eqref{eq:mE}, the operator would be blind to the soft logarithms and only tells how the hard logarithms $\ln(\frac{\Lambda}{\mu})$ are related to one another. Whereas in the EFT where the quarks fields are absent, the massive RG equation instead reveals how the soft logarithms are organized. They follow a clear pattern according to a power counting of the degree of divergence appearing in the diagrammatic expansion. Using the EFT RG operator of \eqref{eq:RGeqHTL} on the generic form of the soft pressure
\begin{equation}\label{eq:PHTL_generic}
    \mathcal{P}^{\rm HTL}(m_E,\as) = m_E^4 \sum_{p=1}^{\infty} \as^{p-1}\sum_{l=0}^{p} a_{p,l} 
    \ln^{p-l} \left(\frac{m_E}{\Lambda}\right),
\end{equation}
exhibits\cite{Fernandez:2021jfr}, after simple algebra, recurrence relations between the different sets of soft logarithms. The simplest is the leading-logarithm (LL) set whose relation reads
\begin{equation}
    -p\, a_{p,0}=\left(4\,\gamma_0^g+2\, b_0^g(p-2)\right)\, a_{p-1,0},
\end{equation}
stemming from $a_{1,0}$ that corresponds to the coefficient of $\ln(\frac{m_E}{\Lambda})$ in Eq.(\ref{eq:PEFT_NLO}).
By direct calculation it has been found that $\gamma_0^g=b_0^g$ \cite{HTLpt}  such that this relation simplifies to a geometric series
\begin{equation}
    a_{p,0}=-2b_0^g\,a_{p-1,0},
\end{equation}
 which in turn, easily resums the LL set in Eq.\eqref{eq:PHTL_generic} as 
\begin{equation}
\begin{aligned}
 \mathcal{P}^{\rm HTL,\, LL}_{l=0}(m_E,\as)=
 m_E^4\, a_{1,0} \frac{\ln\left(\frac{m_E}{\Lambda}\right)}{1+2b_0^g\,\as \ln\left(\frac{m_E}{\Lambda}\right)}.
\end{aligned}
\end{equation}
From here, we match to the full QCD by incorporating the hard degrees of freedom similarly to Eq.\eqref{eq:Matching}. This procedure reproduces the standard massless pressure given in \eqref{eq:PQCD_massless} but also includes the resummed soft LL series. In Fig.\ref{fig:ResumLL}, we compare the NNLO weak-coupling expansion pressure versus its resummed version. There is a net $\sim 20\%$ reduction in uncertainty on the renormalization scale following the resummation procedure providing improved control towards lower $\mu_B$ values.
Note finally that at NLO ${\cal O}(m_E^4 \alpha_s)$ within the EFT, 
our results\cite{Fernandez:2021jfr} combined with those in 
\cite{Gorda:2021znl} imply that the EFT counterterms are local, thus showing the renormalizability of
the $T = 0$ HTL pressure at this order, i.e. N$^3$LO in $\alpha_s^3$.
We refer to \cite{Fernandez:2021jfr} for a more detailed description and considerations
on the next-to-leading soft logarithm resummation.

\section{Renormalization Group optimized perturbation: cold dense QCD}
We now review an approach more related to Screened Perturbation Theory (SPT)\cite{spt1,spt3} 
or more precisely HTLpt\cite{HTLpt,HTLpt2g} within the HTL framework for QCD. 
The latter has been successful at high temperature and zero to small (with respect to temperature) baryonic densities. 
Yet, it has been observed that from NLO to NNLO HTLpt\cite{HTLptDense2L,HTLptDense3L}, the residual scale dependence worsens. 
As mentioned in the introduction, RGOPT\cite{RGOPTals,rgopt_phi4} is essentially a RG-improved version of SPT or HTLpt, indeed, drastically reducing the residual renormalization scale dependence\cite{rgopt_hot} 
as compared to HTLpt.
Moreover, HTLpt has been less explored in the strict zero temperature and high baryonic density limit. 
This may be partly due to the fact that the cold quark matter weak-coupling expansion, involving powers of $\ln\alpha_S$, 
displays better convergence than its counterpart at high temperature,
involving powers of $(\alpha_s)^{1/2}$. Thus, the need for extra resummations (beyond the 
one necessary for IR soft gluon modes) may appear less crucial.
However, even though the $T = 0, \mu_B \ne 0$ NNLO pressure is formally (perturbatively) RG invariant, 
it still exhibits a rather
sizeable scale dependence at intermediate $\mu_B$, specially in the massive quark 
case\cite{Kurkela:2009gj}.  
Thus an application of RGOPT to cold quark matter appears motivated,
and beyond its first application at NLO\cite{rgopt_cold} it was further extended at 
NNLO\cite{RGOPT_ColdDense_NNLO}. 

\subsection{RGOPT procedure}
Observing that in the massive quark case\cite{Kurkela:2009gj}, the quite sizeable residual scale dependence mainly originates from contributions of the hard sector (numerically dominating the NNLO contributions), we consider RGOPT resummation only in the quark
sector to simplify\footnote{One could in principle also apply RGOPT to the gluon effective HTL mass operator \eqref{eq:Lagrangien_HTL}, as is done in HTLpt, but it requires presently missing higher order RG contributions 
in HTL EFT and is left for future work.}. 
Let us first briefly review the OPT approach: in the QCD quark sector, it reads, after standard QCD coupling and quark mass renormalization,
\begin{equation}\label{eq:OPTdelta}
    \mathcal{L}_{\rm  QCD}(g^2)\to \mathcal{L}_{\rm QCD}(g^2\,\delta)+\left(1-\delta\right)\,m\, \bar q q,
\end{equation}
where $m$ is an arbitrary trial mass, subsequently determined by a variational prescription. 
The parameter $0\le \delta \le 1$ interpolates between the free massive and interacting massless theory respectively, and $\delta\,m$ is treated as an interaction.
The physical quantity (here the pressure) is then expanded in powers of $\delta$ at the relevant perturbative order in $g^2$, taking afterward the massless limit $\delta\to 1$. Without $\delta$-expansion this procedure would be a tautology; however, truncating the series at finite $\delta$-order leaves a remnant arbitrary $m$-dependence, commonly fixed through requiring stationarity. This produces a self-consistent mass-gap equation incorporating an all-order summation of certain topologies.

Now, the most important differences between the latter OPT and RGOPT are 
threefold\cite{RGOPTals,RGOPT_ColdDense_NNLO}. 
The first is to (re)introduce (prior to the substitutions in \eqref{eq:OPTdelta}) all vacuum contributions, including contributions from the vacuum energy anomalous dimension\cite{vacEn}, $\propto m^4\,\mathbb{1}$, which mixes with the $m \bar q q$ operator. 
Often disregarded in the thermal literature since medium-independent,  
those vacuum contributions actually play a crucial role in RG properties of a massive theory. Moreover, 
the dressed mass $\mt$ as obtained by RGOPT has effectively a non-trivial dependence both on $g$ and in-medium variables. 

The second difference is to modify the linear $\delta$-interpolation in \eqref{eq:OPTdelta} by introducing a
critical parameter ``$a$'', as
\begin{equation}\label{eq:RGOPTdelta}
   m \left(1-\delta\right) \to m \left(1-\delta\right)^a 
\end{equation}
uniquely fixed to $a=\gamma_0/(2b_0)$ by the RG equation imposed at leading order, 
with $b_0, \gamma_0$ the RG coefficients of the coupling and quark mass anomalous dimension.
This remains consistent with standard renormalization and maintains perturbative
RG invariance {\em after} the $\delta$-expansion is performed.

Lastly, the third modification in RGOPT concerns the determination of the fictitious trial mass parameter. What has been used extensively in OPT (see, e.g. \cite{OPT_PMS,OPT_LDE}) is to fix $m$ using a stationarity principle:
\begin{equation}\label{eq:OPTequation}
    \frac{\partial\mathcal{P}_{\rm QCD}}{\partial m}\Big|_{m=\tilde{m}}=0.
\end{equation}
Unfortunately, the number of possible solutions of \eqref{eq:OPTequation} increases with perturbative order, with no guarantee of real-valued ones.  
An alternative prescription at NNLO HTLpt to circumvent this issue is to replace the
non-trivial mass gap from \eqref{eq:OPTequation} by the standard perturbative Debye mass\cite{HTLptDense3L}. 

In contrast, the RGOPT prescription in \eqref{eq:RGOPTdelta} also guarantees that at successive orders,
essentially only one solution $\mt$ satisfies the leading behavior of the parameter's RG flow in the original (massless) theory\cite{RGOPTals}. For QCD, it implies that the solution will satisfy asymptotic freedom. 
Yet, such ``exact'' dressed solution is again not necessarily real-valued. 
To preserve the all-order RG resummation benefits, we keep the exact mass gap, and in general we can circumvent the complex-valuedness issue by using renormalization scheme (RSC) change, as specified below. 

Moreover, beyond LO it appears more compelling to use the RG equation
\begin{equation}\label{eq:RGreduite}
    \left(\Lambda\frac{\partial}{\partial\Lambda}+\beta(g^2)\frac{\partial}{\partial g^2} \right)\mathcal{P}_{\rm QCD}(g^2,m)\Big|_{m=\mt}=0,
\end{equation}
instead of \eqref{eq:OPTequation} to fix the variational parameter $m$. 
Incidentally, for the cold quark matter pressure at NNLO only \eqref{eq:RGreduite} gives a solution perturbatively consistent
with a screening behavior\cite{RGOPT_ColdDense_NNLO}, $\tilde m_{RG} = {\cal O}(g \mu)$ for $g\to 0$, in contrast to \eqref{eq:OPTequation}, 
thus rendering our prescription essentially uniquely defined at NNLO.
Noting that in the \msb scheme the perturbative coefficients of the relevant quark matter pressure lead to complex-valued solutions already at NLO\cite{rgopt_cold}, one can nevertheless exploit the renormalization scheme freedom to slightly shift the perturbative coefficients from the original \msb result to restore a real-valued $\mt$ solution. Concretely, such RSC is realized using
\begin{equation}\label{eq:RSC}
    m\to m\left(1+B_2 g^4\right),
\end{equation}
and the minimal departure from \msb is obtained by solving for $B_2$ 
and $m$ simultaneously the RG Eq.\eqref{eq:RGreduite} 
and the determinant constraint\cite{RGOPTals,RGOPT_ColdDense_NNLO} 
\begin{equation}\label{eq:determinant}
    \left(\frac{\partial f_{\rm RG}}{\partial g^2}\frac{\partial f_{\rm OPT}}{\partial m}- \frac{\partial f_{\rm OPT}}{\partial g^2} \frac{\partial f_{\rm RG}}{\partial m}\right)\Big|_{m=\mt,\ B_2=\tilde{B}_2}=0 .
\end{equation}
The latter expresses ``contact'' of the $f_{\rm RG}$ and $f_{\rm OPT}$ curves, respectively referring to \eqref{eq:RGreduite} and \eqref{eq:OPTequation},
implying real $\mt$ solution.

\subsection{Application of RGOPT at NNLO for cold and dense QCD}
The NNLO pressure of cold and dense QCD for two massless $(u,d)$ quarks and one massive (strange) quark (referred to as $N_f=2+1^*$) was derived first in \cite{Kurkela:2009gj} with the full $m_s$ dependence. 
Now, to apply the RGOPT procedure we need some modifications of the latter results to deal either with arbitrary degenerate quark masses, $N_f=3^*$ (for medium-dressed mass contributions), or for the more
complete setup incorporating both a medium-dressed mass and the physical strange quark 
mass ($N_f=2^*+1^*$). We reproduced all existing results of \cite{Kurkela:2009gj}\footnote{Apart from a correction\cite{RGOPT_ColdDense_NNLO} in one of the mass-fitting function $G_2(m)$,
with numerically very minor discrepancies with respect to \cite{Kurkela:2009gj} for the physical $m_s$ values.} and generalized those to three different masses, for which $N_f=2^*+1^*$ is a specific case. 

Upon applying the previously
described RGOPT procedure from \eqref{eq:RGOPTdelta} to cold and dense $N_f=2^*+1^*$ NNLO pressure for $(m,m,m+m_s)$ (where $m$ is the variational mass parameter and $m_s$ the physical strange quark mass) and $\mu_u=\mu_d=\mu_s$, 
at NNLO, solving the RG and constraint equations, \eqref{eq:RGreduite},\eqref{eq:determinant}, we obtain the duet of real solutions ($\mt$,$\tilde{B}_2$). 
The resulting complete pressure expression $P_{\rm RGOPT}(\tilde m,
\tilde B_2,\alpha_s,\mu)$ being quite involved\cite{RGOPT_ColdDense_NNLO}, it is more convenient to reproduce it through the following compact formula, obtained by a fitting procedure\footnote{Note that in \eqref{eq:Fit_RGOPT_2p1} 
all the highly nontrivial dependence on running 
$\alpha_S(X \mu)$, $\tilde m(X \mu)$, $\tilde m_s(X \mu)$ from the actual calculation
is conveniently embedded in the fitted $X$-dependence.}:
\begin{eqnarray}\label{eq:Fit_RGOPT_2p1}
 && \frac{\mathcal{P}_{\rm RGOPT}^{N_f=2^*+1^*}(\mu,\Lambda=X\, \mu)}
 {\mathcal{P}_{f}(\mu,N_f=3)} =\left(c_1 + c_2  X^{\nu_3}\right) - \frac{ d_1   (3 \Tilde{\mu})^{\alpha_1}X^{\nu_1}}{(3 \Tilde{\mu} - d_2\, X^{-\nu_2}) },\ \ \  
 \Tilde{\mu}=\mu/{\rm GeV}, \nonumber \\
&& c_1=0.766035 , c_2=0.501495 , \alpha_1=0.996305, d_1=0.402405, \nonumber\\
&& d_2=0.974897, \nu_1=0.410395,\ \nu_2=0.631054,\ \nu_3=0.366230.
\end{eqnarray}
\begin{figure}[h!]
	\centering
    \includegraphics[scale=1.2]{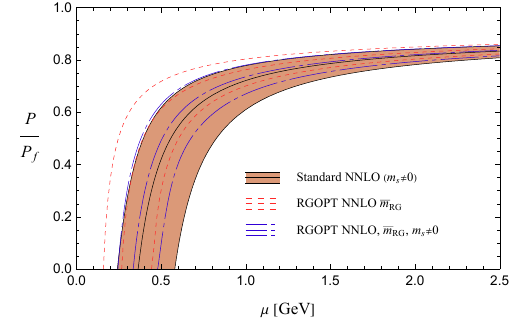}
    \caption{$N_f=3^*$ (red dashed) and $N_f=2^*+1^*$ (blue dot dashed) RGOPT pressures compared to the NNLO pQCD  $N_f=2+1^*$ pressure for 
    $\Lambda= 2\mu X $ with $X \in \left[\frac{1}{2},1,2\right] $.
     \label{Fig:PressureNNLOms}}
\end{figure}
The resulting pressures for $N_f=3^*$ and $N_f=2^*+1^*$ given by \eqref{eq:Fit_RGOPT_2p1} are plotted in Fig.~\ref{Fig:PressureNNLOms} against the standard NNLO weak-coupling expansion result in \cite{Kurkela:2009gj}. Our result fits very well inside the uncertainty range of standard pQCD result but displaying a net improvement with respect to the latter upon variation of the renormalization scale.

\section{Conclusions}
In the massless quark approximation, upon exploiting RG properties of HTL EFT we have obtained an all-order resummation of the soft logarithm contributions to the cold and dense QCD pressure, usefully complementing recent progress
in direct higher order (N$^3$LO) calculations\cite{Gorda:2021kme,Gorda:2023mkk,Karkkainen:2025nkz,Navarrete:2024zgz}
to reduce theoretical uncertainties from residual renormalization scale dependence.
In the more involved massive quark case, the RGOPT modification implied from \eqref{eq:RGOPTdelta} provides a different kind of resummation within a ``variationally improved'' perturbative expansion. At NNLO, our prescription using RG Eq.\eqref{eq:RGreduite} together with \eqref{eq:RSC},\eqref{eq:determinant} 
embeds higher order RG cancellations. 
For both the $N_f = 3^*$ (i.e. $m_s=0$) approximation, or for 
$N_f = 2^* +1^*$ ($m_s \ne 0$), the
residual scale dependence is accordingly reduced as compared to the standard NNLO pressure.
However, these RG improvements appear not to be as efficient as for hot QCD\cite{rgopt_hot}, which may be due to the comparatively better convergence of standard weak-coupling expansion for cold quark matter.
Nevertheless, these different RG resummation improvements
should usefully reduce the present
pQCD uncertainties in the intermediate $\mu_B$ range
relevant for compact stars.

\section*{Acknowledgments}
L.F. has been supported in part by the Research Council of Finland grants Nos. 353772 and 354533.

\end{document}